\documentclass{article}
\usepackage{amsmath,amssymb,amsfonts}

\def\rank{\mathrm{rank}}

\newtheorem{theorem}{Theorem}
\newtheorem{proposition}{Proposition}
\newtheorem{corollary}{Corollary}
\newtheorem{lemma}{Lemma}

\newtheorem{example}{Example}
\newtheorem{definition}{Definition}

\author{Vladimir Dragovi\'c\\{\small Mathematical Institute SANU,
Belgrade, Serbia and Montenegro}\\{\small{\tt
vladad@mi.sanu.ac.yu}}\\and\\
Milena Radnovi\'c\\{\small Mathematical Institute SANU, Belgrade,
Serbia and Montenegro}\\{\small{\tt milena@mi.sanu.ac.yu}}}

\title{Cayley-Type Conditions for Billiards within $k$ Quadrics in $\mathbb{R}^d$}

\date{}

\begin{document}

\maketitle

\begin{abstract}
The notions of reflection from outside, reflection from inside and
signature of a billiard trajectory within a quadric are
introduced. Cayley-type conditions for periodical trajectories for
the billiard in the region bounded by $k$ quadrics in $\mathbb
R^d$ and for the billiard ordered game within $k$ ellipsoids in
$\mathbb R^d$ are derived. In a limit, the condition describing
periodic trajectories of billiard systems on a quadric in
$\mathbb R^d$ is obtained.
\end{abstract}

\section{Introduction}

We study periodic trajectories of the following well-known
integrable mechanical system: motion of a free particle within an
ellipsoid in the Euclidean space of any dimension $d$. On the
boundary, the particle respects the billiard law. To be more
precise, let us mention some basic notions.

Let $(Q,g)$ be a $d$--dimensional Riemannian manifold and let
$D\subset Q$ be a domain with a piecewise smooth boundary
$\mathcal B$. Let $\pi: T^*Q \to Q$ be a natural projection and
let $g^{-1}$ be the contravariant metric on the cotangent bundle.

Consider the {\it reflection mapping} $ r: \pi^{-1} \mathcal B \to
\pi^{-1} \mathcal B, \ p_- \mapsto p_+ , $ which associates the
covector $p_+\in T^*_x Q$, $x\in \mathcal B$ to a covector $p_-\in
T^*_x Q$ such that  the  following conditions hold: $ \vert p_+
\vert =\vert p_- \vert, \ p_+-p_- \bot \mathcal B. $

A {\it billiard} in $D$ is a dynamical system with the phase space
$M=T^*D$ whose trajectories are geodesics given by the Hamiltonian
$ H(p,x)=\frac12 g^{-1}_x(p,p), $ reflected at points $x\in
\mathcal B$ according to the billiard law: $r(p_-)=p_+$. Here
$p_-$ and $p_+$ denote the momenta before and after the
reflection.

Integrability of the billiard system within quadrics is related to
classical geometrical properties: the Chasles, Poncelet and Cayley
theorems. From the Chasles theorem [1] every line in $\mathbb R^d$
is tangent to $d-1$ quadrics confocal to the boundary; all
segments of one billiard trajectory are tangent to the same $d-1$
quadrics [2]. We refer to these $d-1$ quadrics as caustics of the
given trajectory.

From now on, we consider a billiard with a boundary which consists
of the union of $k$ confocal quadrics in $\mathbb R^d$. According
to the generalized Poncelet theorem [3], with $k, d$ arbitrary,
{\it there exists a closed trajectory with $d-1$ given confocal
caustics if and only if infinitely many such trajectories exist,
and all of them have the same period.} The periodicity of a
billiard trajectory depends on its caustic surfaces. An important
question is to find an analytical connection between them and the
corresponding period.

In [4], Cayley found the analytical condition for caustic conics
in the Euclidean plane case ($d=2$) with $k=1$ conic as a
boundary. The classical and algebro-geometric proofs of  Cayley's
theorem can be found in Lebesgue's book [5] and the paper [6],
respectively. Moreover, in [5] the complete Poncelet theorem for
billiard systems in a plane ($d=2$) within $k$ conics, with $k$
arbitrary, was proved (see also [7]).

The generalisation of Cayley's condition for $k=1$ is established
by the authors for any dimension $d$ [8], by use of the
Veselov-Moser discrete $L-A$ pair [9].

The main goal of this paper is to give Cayley-type conditions
describing periodic trajectories of the billiard in the region
bounded by $k$ confocal quadrics in $\mathbb R^d$ and of the
billiard ordered game within $k$ ellipsoids in $\mathbb R^d$, for
$k, d$ arbitrary. The importance of these questions was underlined
several times by experts; let us mention Arnol'd (see [1, 10]),
for example, in connection with applications in laser technology.
In a limit case, we derive analytic conditions for periodic
billiard trajectories on a quadric in $\mathbb R^d$ bounded by any
finite number of quadrics, solving in this way a problem
explicitely posed by Abenda and Fedorov [11].

\section{Planar case: $d=2$, $k$ arbitrary}

The derivation of Cayley-type conditions for the billiard in a
plane within $k$ conics can be done following Lebesgue. In [5], he
considered polygons inscribed in a conic $\Gamma$, whose sides are
tangent to $\Gamma_1,\dots,\Gamma_k$, where $\Gamma,
\Gamma_1,\dots,\Gamma_k$ all belong to a pencil of conics. In the
dual plane, such polygons correspond to billiard trajectories
having caustic $\Gamma^*$ with bounces on
$\Gamma_1^*,\dots,\Gamma_k^*$. The main object of Lebesgue's
analysis in [5] was the cubic Cayley curve, which parametrizes
contact points of tangents drawn from a given point to all conics
of the pencil.

We summarize Lebesgue's results as follows. Let $C$ and $\Gamma$
be conics of a pencil $\mathcal F$ and $\Delta(x)$ be the
discriminant of the conic $C+x\Gamma=0$. If
$\lambda_1,\dots,\lambda_k$ denote parameters corresponding to
$\Gamma_1,\dots,\Gamma_k$, respectively, then the existence of the
Poncelet polygon is equivalent to
$$
\det\left( \begin{array}{ccccccccc} 1 & \lambda_1 & \lambda_1^2 &
\dots & \lambda_1^p & \sqrt{\Delta(\lambda_1)} &
\lambda_1\sqrt{\Delta(\lambda_1)} & \dots &
\lambda_1^{p-2}\sqrt{\Delta(\lambda_1)}\\
\dots\\
\dots\\
1 & \lambda_k & \lambda_k^2 & \dots & \lambda_k^p &
\sqrt{\Delta(\lambda_k)} & \lambda_k\sqrt{\Delta(\lambda_k)} &
\dots & \lambda_k^{p-2}\sqrt{\Delta(\lambda_k)}
\end{array}\right)=0
$$
for $k=2p$
$$
\det\left(\begin{array}{ccccccccc}
 1 & \lambda_1 & \lambda_1^2 & \dots &
\lambda_1^p & \sqrt{\Delta(\lambda_1)} &
\lambda_1\sqrt{\Delta(\lambda_1)} & \dots &
\lambda_1^{p-1}\sqrt{\Delta(\lambda_1)}\\
\dots\\
\dots\\
1 & \lambda_k & \lambda_k^2 & \dots & \lambda_k^p &
\sqrt{\Delta(\lambda_k)} & \lambda_k\sqrt{\Delta(\lambda_k)} &
\dots & \lambda_k^{p-1}\sqrt{\Delta(\lambda_k)}
\end{array}\right)=0
$$
for $k=2p+1$.

The case with two ellipses, when the billiard trajectory is placed
between them and the particle bounces from one to the other of
them alternately, is of special interest.

\begin{corollary} The condition for existence of a $2m$-periodic
billiard trajectory which bounces exactly $m$ times to the ellipse
$\Gamma_1^*=C^*$ and $m$ times to $\Gamma_2^*=(C+\gamma\Gamma)^*$,
having $\Gamma^*$ for the caustic, is
$$
\det\left(\begin{array}{cccc}
f_0(0) & f_1(0) &\dots & f_{2m-1}(0)\\
f_0'(0) & f_1'(0) &\dots & f_{2m-1}'(0)\\
\dots\\
f_0^{(m-1)}(0) & f_1^{(m-1)}(0) &\dots & f_{2m-1}^{(m-1)}(0)\\
f_0(\gamma) & f_1(\gamma) &\dots & f_{2m-1}(\gamma)\\
f_0'(\gamma) & f_1'(\gamma) &\dots & f_{2m-1}'(\gamma)\\
\dots\\
f_0^{(m-1)}(\gamma) & f_1^{(m-1)}(\gamma) &\dots &
f_{2m-1}^{(m-1)}(\gamma)
\end{array}\right)=0
$$
where $f_j=x^j$ $(0\le j\le m)$, $f_{m+i}=x^{i-1}\sqrt{\Delta(x)}$
$(1\le i\le m-1)$.
\end{corollary}

\medskip

We consider a simple example with four bounces on each of the two
conics.

\begin{example} The condition on a billard trajectory placed
between ellipses $\Gamma_1^*$ and $\Gamma_2^*$, to be closed after
four alternating bounces to each of them is
$$
\det X = 0
$$
where the elements of the $3\times 3$ matrix $X$ are
$$
\aligned
X_{11}&= -4B_0+B_1\gamma+4C_0+3C_1\gamma+2C_2\gamma^2+C_3\gamma^3 \\
X_{12}&=-3B_0+B_1\gamma+3C_0+2C_1\gamma+C_2\gamma^2 \\
X_{13}&=-2B_0+B_1\gamma+2C_0+C_1\gamma \\
X_{21}&=-6B_0+B_2\gamma^2+6C_0+6C_1\gamma+4C_2\gamma^2+3C_3\gamma^3 \\
X_{22}&=-6B_0+B_1\gamma+B_2\gamma^2+6C_0+4C_1\gamma+3C_2\gamma^2 \\
X_{23}&=-5B_0+2B_1\gamma+B_2\gamma^2+5C_0+3C_1\gamma \\
X_{31}&=-4B_0+B_3\gamma^3+4C_0+4C_1\gamma+4C_2\gamma^2+3C_3\gamma^3 \\
X_{32}&=-4B_0+B_2\gamma^2+B_3\gamma^3+4C_0+4C_1\gamma+3C_2\gamma^2 \\
X_{33}&=-4B_0+B_1\gamma+B_2\gamma^2+B_3\gamma^3+4C_0+3C_1\gamma
\endaligned
$$
with $C_i, B_i$ being coefficients in the Taylor expansions around
$x=0$ and $x=\gamma$, respectively
$$
\aligned
\sqrt{\Delta(x)}&=C_0+C_1x+C_2x^2+\dots\\
\sqrt{\Delta(x)}&=B_0+B_1(x-\gamma)+B_2(x-\gamma)^2+\dots.
\endaligned
$$
\end{example}

\section{Periodic billiard trajectories inside $k$ confocal
quadrics in $\mathbb R^d$}

The complete Poncelet theorem (CPT) was generalized to the case
$d=3$ by Darboux in [12] in 1870. Higher-dimensional
generalizations of CPT were obtained quite recently in [3]. The
main result of the present paper is the Cayley-type condition for
generalized CPT for $d\ge 3$, although obtained results can be
applied immediately in the case $d=2$.

\smallskip

Consider an ellipsoid in ${\mathbb R}^d$
$$
\frac {x_1^2}{a_1}+\dots + \frac {x_d^2}{a_d}=1\qquad
a_1>\dotsb>a_d>0
$$
and a related system of Jacobian elliptic coordinates
$(\lambda_1,\dots, \lambda_d)$ ordered by the condition
$$
 \lambda_1>\lambda_2>\dotsb> \lambda_d.
$$
Any quadric from the corresponding confocal family is given by
\begin{equation}
Q_{\lambda}:\ \frac {x_1^2}{a_1-\lambda}+\dots + \frac
{x_d^2}{a_d-\lambda}=1.
\end{equation}

\begin{lemma} Suppose a line $\ell$ is tangent to quadrics
$Q_{\alpha_1},\dots,Q_{\alpha_{d-1}}$ from the family $(1)$. Then
Jacobian coordinates $(\lambda_1,\dots, \lambda_d)$ of any point
on $\ell$ satisfy the inequalities $\mathcal P(\lambda_s)\ge 0$,
$s=1,\dots,d$, where
$$
\mathcal
P(x)=(a_1-x)\dots(a_d-x)(\alpha_1-x)\dots(\alpha_{d-1}-x).
$$
\end{lemma}

\noindent{\it Proof.} Follows from [13]. \hfill $\Box$

\medskip

Suppose that a bounded domain $\Omega\subset\mathbb R^d$ is given
such that its boundary $\partial\Omega$ lies in the union of
several quadrics from the family (1). Then, in elliptic
coordinates, $\Omega$ is given by:
$$
\beta_1'\le\lambda_1\le\beta_1'', \quad\dots,\quad
\beta_d'\le\lambda_d\le\beta_d'',
$$
where $a_{s+1}\le\beta_s'<\beta_s''\le a_s$ for $1\le s\le d-1$
and $-\infty<\beta_d'<\beta_d''\le a_d$.

Consider a billiard system within $\Omega$ and let
$Q_{\alpha_1},\dots,Q_{\alpha_{d-1}}$ be caustics of one of its
trajectories. For any $s=1,\dots, d$, the set $\Lambda_s$ of all
values taken by the coordinate $\lambda_s$ on the trajectory is,
according to lemma 1, included in $\Lambda_s'=\{\,
\lambda\in[\beta_s',\beta_s'']\, :\, \mathcal P(\lambda)\ge0\,\}$.
By [14], the set $\Lambda_s'$ is a closed interval and coincides
with $\Lambda_s$. Denote
$[\gamma_s',\gamma_s'']:=\Lambda_s=\Lambda_s'$.

Note that the trajectory touches quadrics of any pair
$Q_{\gamma_s'}, Q_{\gamma_s''}$ alternately. Thus, any periodic
trajectory has the same number of intersection points with each of
them.

\begin{theorem} A trajectory of the billiard system
within $\Omega$ with caustics $Q_{\alpha_1}$, \dots,
$Q_{\alpha_{d-1}}$ is periodic with exactly $n_s$ points at
$Q_{\gamma_s'}$ and $n_s$ points at $Q_{\gamma_s''}$ $(1\le s\le
d)$ if and only if
\begin{equation}
\sum_{s=1}^d n_s\left(\mathcal A(P_{\gamma_s'}) -\mathcal
A(P_{\gamma_s''})\right)=0
\end{equation}
on the Jacobian of the curve
$$
\Gamma \ :\ y^2=\mathcal P(x):=
(a_1-x)\cdots(a_d-x)(\alpha_1-x)\cdots(\alpha_{d-1}-x).
$$
Here $\mathcal A$ denotes the Abel-Jacobi map, where
$P_{\gamma_s'}$, $P_{\gamma_s''}$ are points on $\Gamma$ with
coordinates $P_{\gamma_s'}=\left(\gamma_s', (-1)^s \sqrt {\mathcal
P(\gamma_s')}\right)$, $P_{\gamma_s''}=\left(\gamma_s'', (-1)^s
\sqrt {\mathcal P(\gamma_s'')}\right)$.
\end{theorem}

\noindent{\it Proof.} Following Jacobi [15], let us consider the
integrals
$$
\sum_{s=1}^d\int{d\lambda_s\over \sqrt{\mathcal P(\lambda_s)}},
\quad \sum_{s=1}^d\int{\lambda_s d\lambda_s\over \sqrt{\mathcal
P(\lambda_s)}}, \quad \dots, \quad
\sum_{s=1}^d\int{\lambda_s^{d-1} d\lambda_s\over \sqrt{\mathcal
P(\lambda_s)}}
$$
over the polygonal line $A_1A_2\dots A_{k+1}$, which represents a
billiard trajectory, where $k=2(n_1+\dots+n_d)$. The last integral
is equal to the total length of the polygonal line, while the
others are equal to zero. Considering the behaviour of elliptic
coordinates along each segment of the trajectory, we calculate
values of the integrals and obtain that the condition
$A_{k+1}=A_1$ is equivalent to (2). \hfill$\Box$

\medskip

Our  next step is to introduce a notion of bounces `from outside'
and `from inside'. More precisely, let us consider an ellipsoid
$Q_{\lambda}$ from the confocal family (1) such that $\lambda \in
(a_{s+1}, a_s)$ for some $s\in \{1,\dots, d\}$, where
$a_{d+1}=-\infty$.

Observe that along a billiard ray which reflects at $Q_{\lambda}$,
the elliptic coordinate $\lambda_i$ has a local extremum at the
point of reflection.

\begin{definition} A ray reflects {\rm from outside} at
the quadric $Q_{\lambda}$ if the reflection point is a local
maximum of the Jacobian coordinate $\lambda_s$, and it reflects
{\rm from inside} if the reflection point is a local minimum of
the coordinate $\lambda_s$.
\end{definition}

Let us remark that in the case when $Q_{\lambda}$ is an ellipsoid,
the notions introduced in definition 1 coincide with the usual
ones.

Assume now a $k$-tuple of confocal quadrics $Q_{\beta_1},\dots
,Q_{\beta_k}$ from the confocal pencil (1) is given. We consider a
billiard system with trajectories having bounces at
$Q_{\beta_1},\dots, Q_{\beta_k}$ respectively. Such a trajectory
has $d-1$ caustics from the same family (1). We additionally
assign to each trajectory {\it the signature}
$\sigma=(i_1,\dots,i_k)$ by the following rule:
$$
\aligned
i_s &= +1 \qquad \text {if the reflection at}\; Q_{\beta_s}\; \text {is from inside}\\
i_s &= -1 \qquad \text {if the reflection at}\; Q_{\beta_s}\; \text{is from outside}.\\
\endaligned
$$

Suppose $Q_{\beta_1}$, \dots, $Q_{\beta_k}$ are ellipsoids and
consider a {\it billiard ordered game} with signature
$\sigma=(i_1,\dots,i_k)$. In order that trajectories of such a
game stay bounded, the following condition has to be satisfied:
$$
i_s=-1\quad \Rightarrow\quad i_{s+1}=i_{s-1}=1 \;\;\;\;
\text{and}\;\;\;\; \beta_{s+1}<\beta_s,\ \beta_{s-1}<\beta_s.
$$
(Here, we identify indices 0 and $k+1$ with $k$ and 1,
respectively.)

\begin{theorem} Given a billiard ordered game within $k$
ellipsoids $Q_{\beta_1},\dots ,Q_{\beta_k}$ with signature
$\sigma=(i_1,\dots,i_k)$. Its trajectory with caustics
$Q_{\alpha_1},\dots ,Q_{\alpha_{d-1}}$ is $k$-periodic if and only
if
$$
\sum_{s=1}^k i_s\bigl(\mathcal A(P_{\beta_s})-\mathcal
A(P_{\alpha})\bigr)
$$
is equal to a sum of several expressions of the form $\mathcal
A(P_{\alpha_p})-\mathcal A(P_{\alpha_{p'}})$ on the Jacobian of
the curve $\Gamma\, :\, y^2=\mathcal P(x),$ where
$P_{\beta_s}=\left(\beta_s,+\sqrt {\mathcal P(\beta_s)}\right)$,
$\alpha =\min\{a_d, \alpha_1,\dots,\alpha_{d-1}\}$ and
$Q_{\alpha_p}$, $Q_{\alpha_{p'}}$ are pairs of caustics of the
same type.
\end{theorem}

When $Q_{\beta_1}=\dotsb=Q_{\beta_k}$ and $i_1=\dotsb=i_k=1$ we
obtain the Cayley-type condition for billiard motion inside an
ellipsoid in $\mathbb R^d$. Such periodic trajectories were
described in [8] using a different technique, based on a
Veselov-Moser discrete Lax representation.

We are going to treat in more detail the case of billiard motion
between two ellipsoids.

\begin{proposition} The condition that there exists a closed
billiard trajectory between two ellipsoids $Q_{\beta_1}$ and
$Q_{\beta_2}$, which bounces exactly $m$ times to each of them,
with caustics $Q_{\alpha_1},\dots,Q_{\alpha_{d-1}}$, is
$$
\rank\left(\begin{array}{cccc} f_1'(P_{\beta_2}) &
f_2'(P_{\beta_2}) & \dots &
f_{m-d+1}'(P_{\beta_2}) \\
f_1''(P_{\beta_2}) & f_2''(P_{\beta_2}) & \dots &
f_{m-d+1}''(P_{\beta_2}) \\
\dots\\
\dots\\
f_1^{(m-1)}(P_{\beta_2}) & f_2^{(m-1)}(P_{\beta_2}) & \dots &
f_{m-d+1}^{(m-1)}(P_{\beta_2})
\end{array}\right)<m-d+1.
$$
Here
$$
f_j= {y-B_0-B_1(x-\beta_1)-\dots-B_{d+j-2}(x-\beta_1)^{d+j-2}
\over x^{d+j-1}} \qquad 1\le j\le m-d+1
$$
and $y=B_0+B_1(x-\beta_1)+\dots$ is the Taylor expansion around
the point symmetric to $P_{\beta_1}$ with respect to the
hyperelliptic involution of the curve $\Gamma$. {\rm (All notation
is as in theorem 2.)}
\end{proposition}

\section{Periodic trajectories of billiards on quadrics in
$\mathbb R^d$}

In [16] the billiard systems on a quadric $\mathcal E$ in $\mathbb
R^d$
$$
\frac {x_1^2}{a_1}+\dots + \frac {x_d^2}{a_d}=1,\quad
a_1>\dotsb>a_d,
$$
are defined as limits of corresponding billiards within $\mathcal
E$, when one of the caustics tends to $\mathcal E$. The boundary
of such a billiard consists of the intersection of $\mathcal E$
with certain confocal quadrics $Q_{\beta_1}, \dots ,Q_{\beta_k}$.
The question of description of periodic trajectories of these
systems was formulated as an open problem by Abenda and Fedorov
[11].

By applying the limit procedure, from theorem 1, we obtain the
following:

\begin{theorem} A trajectory of the billiard system
constrained to the ellipsoid $\mathcal E$ within $\Omega:$
$\beta_1'\le\lambda_1\le\beta_1''$, \dots,
$\beta_{d-1}'\le\lambda_{d-1}\le\beta_{d-1}''$, with caustics
$Q_{\alpha_1}$, \dots, $Q_{\alpha_{d-2}}$, is periodic with
exactly $n_s$ bounces at each of quadrics $Q_{\gamma_s'}$,
$Q_{\gamma_s''}$ $(1\le s\le d-2)$ if and only if
$$
\sum_{s=1}^{d-1} n_s\bigl(\bar{\mathcal A}(P_{\gamma_s'})
-\bar{\mathcal A}(P_{\gamma_s''})\bigr)=0
$$
on the Jacobian of the curve
$$
\Gamma_1 \ :\ y^2=\mathcal
P_1(x):=-x(a_1-x)\cdots(a_d-x)(\alpha_1-x)\cdots(\alpha_{d-2}-x).
$$
Here $P_{\gamma_s'}$, $P_{\gamma_s''}$ are the points on
$\Gamma_1$ with coordinates
$P_{\gamma_s'}=\left(\gamma_s',(-1)^s\sqrt {\mathcal
P_1(\gamma_s')}\right)$,
$P_{\beta_s''}=\left(\gamma_s'',(-1)^s\sqrt {\mathcal
P_1(\gamma_s'')}\right)$, with $[\gamma_s', \gamma_s'']=\{\,
\lambda\in[\beta_s',\beta_s'']\, :\, \mathcal
P_1(\lambda)\ge0\,\}$, $1\le s\le d-2$, and $\bar{\mathcal
A}(P)=(0, \int_0^P \frac{x dx}y,\int_0^P \frac{x^2 dx}y, \dots,
\int_0^P \frac{x^{d-2} dx}y)$.
\end{theorem}

In the same way as in the previous section, a billiard ordered
game constrained to the ellipsoid $\mathcal E$ within given
quadrics $Q_{\beta_1},\dots ,Q_{\beta_k}$ of the same type can be
defined. The only difference is that now the signature
$\sigma=(i_1,\dots,i_k)$ can be given arbitrarily, since
trajectories are bounded, lying on the compact hypersurface
$\mathcal E$. Denote by $Q_{\alpha_1}$, \dots, $Q_{\alpha_{d-2}}$
the caustics of a given trajectory of the game. Since quadrics
$Q_{\beta_1},\dots ,Q_{\beta_k}$ are all of the same type, there
exist $\mu',\mu''$ in the set
$S=\{a_1,\dots,a_d,\alpha_1,\dots,\alpha_{d-2}\}$ such that
$\beta_1,\dots,\beta_k\in[\mu',\mu'']$ and $(\mu',\mu'')\cap S$ is
empty.

Associate with the game the following divisors on the curve
$\Gamma_1$:
$$
\mathcal D_s= \begin{cases}
P_{\mu''} & \text{if}\; i_s=i_{s+1}=1\\
0 & \text{if}\; i_s=-i_{s+1}=1, \beta_s<\beta_{s+1}\; \text{or}\;
i_s=-i_{s+1}=-1, \beta_s>\beta_{s+1}\\
P_{\mu''}-P_{\mu'} & \text{if}\; i_s=-i_{s+1}=1,
\beta_s>\beta_{s+1}\\
P_{\mu'}-P_{\mu''} & \text{if}\; i_s=-i_{s+1}=-1, \beta_s<\beta_{s+1}\\
P_{\mu'} & \text{if}\; i_s=i_{s+1}=-1,
\end{cases}
$$
where $P_{\mu'}$ and $P_{\mu''}$ are its branching points with
coordinates $(\mu',0)$ and $(\mu'',0)$, respectively.

\begin{theorem} Given a billiard ordered game constrained to
$\mathcal E$ within quadrics $Q_{\beta_1}$,\dots, $Q_{\beta_k}$
with signature $\sigma=(i_1,\dots,i_k)$. Its trajectory with
caustics $Q_{\alpha_1}$, \dots, $Q_{\alpha_{d-2}}$ is $k$-periodic
if and only if
$$
\sum_{s=1}^k i_s\bigl(\bar{\mathcal A}(P_{\beta_s})-\bar{\mathcal
A}(\mathcal D_s)\bigr)
$$
is equal to a sum of several expressions of the form
$\bar{\mathcal A}(P_{\alpha_p})-\bar{\mathcal A}(P_{\alpha_{p'}})$
on the Jacobian of the curve $ \Gamma_1 : y^2=\mathcal P_1(x), $
where $P_{\beta_s}=\left(\beta_s,+\sqrt {\mathcal
P_1(\beta_s)}\right)$ and $Q_{\alpha_p}$, $Q_{\alpha_{p'}}$ are
pairs of caustics of the same type.
\end{theorem}

\begin{proposition} Consider the case $d=3$ and a billiard system
constrained to the ellipsoid $\mathcal E$ with the boundary
$Q_{\gamma}$ and caustic $Q_{\alpha}$, $a_3<\gamma<\alpha<a_2$. A
trajectory is $k$-periodic if:
$$
\rank\left(\begin{array}{cccc}
C_{p+1} & C_{p+2} & \dots & C_{2p-2}\\
C_{p+2} & C_{p+3} & \dots & C_{2p-1}\\
\dots\\
C_{2p} & C_{2p+1} & \dots & C_{3p-3}
\end{array}\right)<p-2
\qquad k=2p
$$
and
$$
\rank\left(\begin{array}{cccc}
C_{p+1} & C_{p+2} & \dots & C_{2p-1}\\
C_{p+2} & C_{p+3} & \dots & C_{2p}\\
\dots\\
C_{2p} & C_{2p+1} & \dots & C_{3p-2}
\end{array}\right)<p-1
\qquad k=2p+1
$$
where
$$
y=C_0+C_1\left(\tilde
x-{1\over\alpha-\gamma}\right)+C_2\left(\tilde
x-{1\over\alpha-\gamma}\right)^2+\dots
$$
is the Taylor expansion with respect to $\tilde
x=\dfrac1{\alpha-x}$ around the point $P_{\gamma}$.
\end{proposition}

\section{Conclusion}

As an important historical remark, we would like to emphasize the
significance of Darboux's contribution to the  study of problems
related to generalized Poncelet theorem. The impression is that
his work in the field (see [12]) is completely unkonown nowadays.
We shall present in another publication a more detailed overview
of Darboux's ideas with comparison to the Lebesgue geometric
approach and applications to separable perturbed problems [17,
18].

\subsection*{Acknowledgements}

The research was partially supported by the Serbian Ministry of
Science and Technology, Project 1643 -- Geometry and Topology of
Manifolds and Integrable Dynamical Systems. The research of one of
the authors (VD) was also partially supported by SISSA and MURST
Project Geometry of Integrable Systems. The authors would like to
thank Professors B Dubrovin, Yu Fedorov and S Abenda for
interesting discussions.

\end{document}